\newcommand{\met}{\mbox{${\rm \not\! E}_{\rm T}$}}
\newcommand{\pT}{p$_{\rm T}$}
\newcommand{\Et}{{\rm E}_{\rm T}}
\newcommand{\ET}{E$_{\rm T}$}
\newcommand{\Pt}{{\rm p}_{\rm T}}
\newcommand{\wenu}{$W \rightarrow e\nu$}
\newcommand{\Z}{{ Z^0}}
\newcommand{\gravitino}{\tilde{G}}

\documentclass[prl,twocolumn,superscriptaddress,showpacs,floatfix]{revtex4}
\usepackage[letterpaper]{geometry}
\usepackage{amsfonts}
\usepackage{graphicx}
\usepackage{psfrag}
\usepackage{times}
\begin{document}\setlength{\unitlength}{1mm}
\bibliographystyle{apsrev}

\preprint{CDF/PUB/EXOTIC/CDFR/5765}
\preprint{EFI 01-53}
\preprint{Version 1.3}

\title{Limits on Extra Dimensions and New Particle Production in the
Exclusive Photon and Missing Energy
Signature in $p\bar p$ Collisions at $\sqrt{s} =$ 1.8 TeV}




\affiliation{Institute of Physics, Academia Sinica, Taipei, Taiwan 11529, 
Republic of China}
\affiliation{Argonne National Laboratory, Argonne, Illinois 60439}
\affiliation{Istituto Nazionale di Fisica Nucleare, University of Bologna,
I-40127 Bologna, Italy}
\affiliation{Brandeis University, Waltham, Massachusetts 02254}
\affiliation{University of California at Davis, Davis, California  95616}
\affiliation{University of California at Los Angeles, Los 
Angeles, California  90024} 
\affiliation{Instituto de Fisica de Cantabria, CSIC-University of Cantabria, 
39005 Santander, Spain}
\affiliation{Carnegie Mellon University, Pittsburgh, PA  15218} 
\affiliation{Enrico Fermi Institute, University of Chicago, Chicago, 
Illinois 60637}
\affiliation{Joint Institute for Nuclear Research, RU-141980 Dubna, Russia}
\affiliation{Duke University, Durham, North Carolina  27708} 
\affiliation{Fermi National Accelerator Laboratory, Batavia, Illinois 
60510}
\affiliation{University of Florida, Gainesville, Florida 32611}
\affiliation{Laboratori Nazionali di Frascati, Istituto Nazionale di Fisica
               Nucleare, I-00044 Frascati, Italy}
\affiliation{University of Geneva, CH-1211 Geneva 4, Switzerland} 
\affiliation{Glasgow University, Glasgow G12 8QQ, United Kingdom}
\affiliation{Harvard University, Cambridge, Massachusetts 02138} 
\affiliation{Hiroshima University, Higashi-Hiroshima 724, Japan}
\affiliation{University of Illinois, Urbana, Illinois 61801}
\affiliation{The Johns Hopkins University, Baltimore, Maryland 21218}
\affiliation{Institut f\"{u}r Experimentelle Kernphysik, 
Universit\"{a}t Karlsruhe, 76128 Karlsruhe, Germany}
\affiliation{Center for High Energy Physics: Kyungpook National
University, Taegu 702-701; Seoul National University, Seoul 151-742; and
SungKyunKwan University, Suwon 440-746; Korea}
\affiliation{High Energy Accelerator Research Organization (KEK), Tsukuba, 
Ibaraki 305, Japan}
\affiliation{Ernest Orlando Lawrence Berkeley National Laboratory, 
Berkeley, California 94720}
\affiliation{Massachusetts Institute of Technology, Cambridge,
Massachusetts  02139} 
\affiliation{University of Michigan, Ann Arbor, Michigan 48109}
\affiliation{Michigan State University, East Lansing, Michigan  48824}
\affiliation{University of New Mexico, Albuquerque, New Mexico 87131} 
\affiliation{Northwestern University, Evanston, Illinois  60208} 
\affiliation{The Ohio State University, Columbus, Ohio  43210}
\affiliation{Osaka City University, Osaka 588, Japan} 
\affiliation{University of Oxford, Oxford OX1 3RH, United Kingdom} 
\affiliation{Universita di Padova, Istituto Nazionale di Fisica 
          Nucleare, Sezione di Padova, I-35131 Padova, Italy}
\affiliation{University of Pennsylvania, Philadelphia, 
        Pennsylvania 19104}
\affiliation{Istituto Nazionale di Fisica Nucleare, University and Scuola
               Normale Superiore of Pisa, I-56100 Pisa, Italy} 
\affiliation{University of Pittsburgh, Pittsburgh, Pennsylvania 15260} 
\affiliation{Purdue University, West Lafayette, Indiana 47907}
\affiliation{University of Rochester, Rochester, New York 14627}
\affiliation{Rockefeller University, New York, New York 10021}
\affiliation{Rutgers University, Piscataway, New Jersey 08855}
\affiliation{Texas A\&M University, College Station, Texas 77843}
\affiliation{Texas Tech University, Lubbock, Texas 79409}
\affiliation{Institute of Particle Physics, University of Toronto, Toronto
M5S 1A7, Canada}
\affiliation{Istituto Nazionale di Fisica Nucleare, University of 
Trieste/Udine, Italy}
\affiliation{University of Tsukuba, Tsukuba, Ibaraki 305, Japan}
\affiliation{Tufts University, Medford, Massachusetts 02155}
\affiliation{Waseda University, Tokyo 169, Japan}
\affiliation{University of Wisconsin, Madison, Wisconsin 53706}
\affiliation{Yale University, New Haven, Connecticut 06520}

\author{D.~Acosta}
\affiliation{University of Florida, Gainesville, Florida 32611}

\author{T.~Affolder}
\affiliation{Ernest Orlando Lawrence Berkeley National Laboratory, 
Berkeley, California 94720}

\author{H.~Akimoto}
\affiliation{Waseda University, Tokyo 169, Japan}

\author{M.~G.~Albrow}
\affiliation{Fermi National Accelerator Laboratory, Batavia, Illinois 
60510}

\author{D.~Ambrose}
\affiliation{University of Pennsylvania, Philadelphia, 
        Pennsylvania 19104}

\author{D.~Amidei}
\affiliation{University of Michigan, Ann Arbor, Michigan 48109}

\author{K.~Anikeev}
\affiliation{Massachusetts Institute of Technology, Cambridge,
Massachusetts  02139} 

\author{J.~Antos}
\affiliation{Institute of Physics, Academia Sinica, Taipei, Taiwan 11529, 
Republic of China}

\author{G.~Apollinari}
\affiliation{Fermi National Accelerator Laboratory, Batavia, Illinois 
60510}

\author{T.~Arisawa}
\affiliation{Waseda University, Tokyo 169, Japan}

% ...


\date{\today}

\begin{abstract}
The exclusive $\gamma$\met{} signal has a small standard model 
cross-section and is thus
a good channel in which to look for signs of new physics.  This signature is 
predicted by models
with a superlight gravitino or with large extra spatial dimensions.  We search for such
signals at the CDF detector at the Tevatron, using 87 pb$^{-1}$ of data at 
$\sqrt{s} = 1.8$ TeV, and extract 95\% C.L.\ limits on these processes.
A limit of 221 GeV is set on the scale $|F|^{1/2}$ in supersymmetry 
models.  For 4, 6, and 8 extra dimensions, limits on the fundamental
mass scale $M_D$ of 549, 581, and 602 GeV, respectively, 
are found.  We also specify a `pseudo-model-independent' method of comparing 
the results to theoretical predictions.
\end{abstract}

\maketitle


Many extensions to the standard model predict the existence of
minimally-interacting particles, such as the gravitino in
supersymmetric models and Kaluza-Klein (KK) modes of the graviton in models
with large compact spatial dimensions \cite{ledsref}.  Such particles cannot be
directly observed in a detector, but their production can be inferred
from a transverse momentum imbalance (or ``missing transverse energy,''
$\met{}$
\cite{etdef}) among the visible particles in a high-energy collision.
Photons can be
emitted in such hard-scattering processes due to the presence of charged
quarks in the $p\bar p$ initial state; many models also predict the 
production of photons from the decay of
final-state particles \cite{Ambrosiano}.
A search for the $\gamma\met$ signature thus explores a wide range of
models and complements searches in the single jet$+\met$ channel 
\cite{castro}.  Here we present the results of a search in the
exclusive $\gamma$\met{} signature, i.e.\ where only a photon and invisible
particles are produced.

The data used for this analysis were collected with the Collider Detector 
at Fermilab (CDF)
during Run 1b of the Tevatron, with an integrated luminosity of 
87 $\pm$ 4 pb$^{-1}$ of $p\bar p$ collisions at $\sqrt{s} = 1.8$ TeV.
The CDF detector has been described in detail elsewhere \cite{detector}; 
subsystems most important to this search are summarized here.
A system of time projection chambers around the beampipe allows the
determination of the event vertex position.  Surrounding these chambers is
the central tracking chamber (CTC), a
cylindrical drift chamber inside a 1.4 T superconducting solenoid, which
is fully efficient for track reconstruction in the pseudorapidity region 
$|\eta| < 1.0$ \cite{cdf_coo}.
The central electromagnetic calorimeter (CEM) covers the region
$|\eta| < 1.1$.  Strip chambers (the CES system) are embedded in the CEM at
the depth of shower maximum to
allow the measurement of the 2-dimensional transverse profile
of electromagnetic showers.
The central hadronic calorimeter
covers the range $|\eta| < 1.3$ and is instrumented with 
time-to-digital converters which associate times to shower signals. 
The calorimeter modules are arranged in ``towers,'' with hadronic modules
behind the electromagnetic modules, in a projective geometry pointing
at the center of the detector.  High energy electromagnetic
showers frequently leak from the
electromagnetic modules into the hadronic modules behind them;
when sufficient leakage occurs timing can be associated with the 
electromagnetic 
shower.
Outside the calorimeters, drift chambers provide muon identification in
the region $|\eta| < 1.0$.

To select events with high-\pT{} photons during data-taking, we use
the CDF 3-level trigger system \cite{trigger}.  Level 1 requires a
central calorimeter tower with $\Et^{EM} > 8$ GeV \cite{cdf_coo}.  The Level 2
system forms clusters of towers and then requires the event to pass an
OR of several triggers, including: a) a very loose trigger requiring only
an electromagnetic cluster \cite{top_evidence} with $\Et^{EM} >$ 50 GeV and
b) a trigger requiring \met{} $>$ 35 GeV.  Level 3 requires that
the photon candidate satisfy \ET{} $>$ 50 GeV and have a CES cluster
within the fiducial region \cite{jeff_prd}.

The offline photon candidate identification (``Photon ID'') criteria 
\cite{toback_prd,ray_prd,jeff_prd} are
a) an electromagnetic cluster 
in the CEM with $|\eta^\gamma| <$ 1 \cite{eta_note}, a ratio 
$\mathrm{E}^{HAD}/\mathrm{E}^{EM}$ less than $0.055 + 0.00045 
\times \mathrm{E}^{SUM}$, a centroid 
within the 
fiducial region of the CES, and shower evolution measured by the CES 
consistent with expectation;
b) no second energetic object in the same CES wire chamber as the cluster;
c) at most one CTC track, and none
with \pT{} $>$ 1 GeV \cite{units}, pointing at the cluster;
d) within a radius of 0.4 in $\eta$-$\phi$
space around the cluster centroid, 
\ET{} (summed over towers excluding those in the photon cluster) $<$ 2 GeV 
and a sum of track
\pT{} $<$ 5 GeV;
e) $\Et^\gamma >$ 55 GeV \cite{correctionsnote}; 
and 
f) an event vertex within 60 cm of the center of the detector along the
beamline.

The selection on missing transverse energy is
\met{} $>$ 45 GeV.  This threshold is lower than the 
$\Et^\gamma$ threshold to keep this requirement fully efficient for signal 
processes, taking into consideration the \met{} resolution and 
the intrinsic parton \pT{} in the $p$ and $\bar p$ initial states.

Backgrounds to the $\gamma$\met{} signal include:
a) $q\bar q \rightarrow Z\gamma \rightarrow \nu\bar\nu\gamma$;
b) cosmic ray muons that undergo bremsstrahlung in the CEM but for which 
no track is found;
c) $W \rightarrow e\nu$ with the electron misidentified as a photon;
d) $W\gamma$ production where the charged lepton in a leptonic $W$
decay is lost;
e) prompt $\gamma\gamma$ production where a photon is lost; and 
f) dijet and photon + jet production.  

To reject cosmic ray muons, we require 
a timing signal in the hadronic calorimeter which is 
in-time with the collision within a window 55 ns wide
for at least one tower 
in the cluster, and no evidence of a muon
in the central muon systems 
within $30^\circ$ in $\phi$ of the photon.  The efficiency of requiring that
timing information be present
rises with $\Et^\gamma$ from 78\% at 55 GeV to over 98\% above 100
GeV.  The efficiency of these two cuts is measured with a sample of 
isolated electrons.

To remove the $W\gamma$ background as well as events in which 
mismeasurement of jet energy produces fake
\met{},
we require 
no jets \cite{top_evidence} with \ET{} $>$ 15 GeV,
no jets with \ET{} $>$ 8 GeV within 0.5 radians in $\phi$ of the photon,
and
no tracks in the event with $\Pt >$ 5 GeV.  

\begin{figure}
\includegraphics[width=\linewidth]{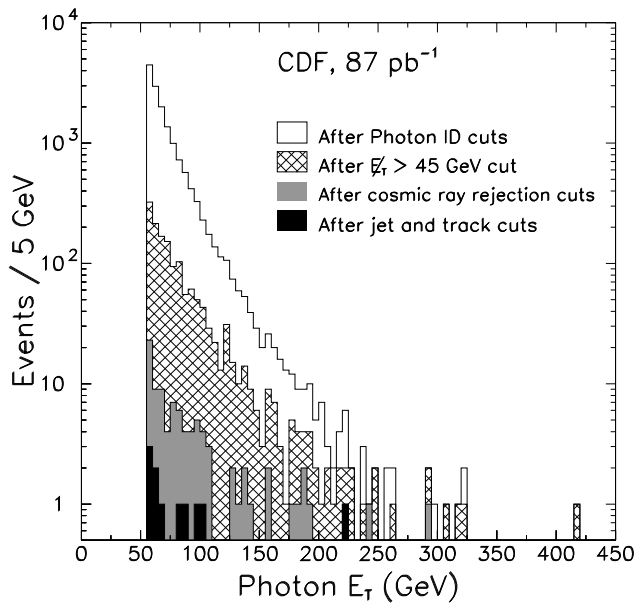}
\caption{Photon \ET{} spectrum for events remaining after 
each stage of cuts.  
}
\label{fig:staged_spectrum}
\end{figure}

\begin{figure}
\includegraphics[width=\linewidth]{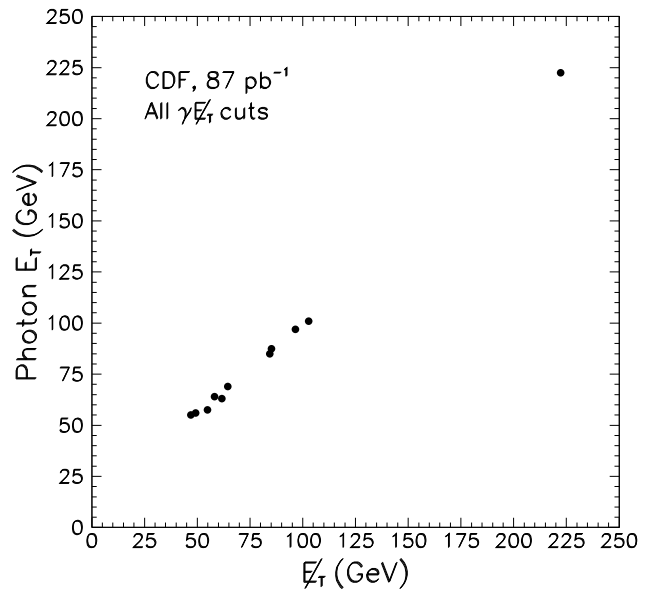}
\caption{Photon \ET{} versus \met{} for the 11 events passing the selection
criteria. The tight correlation of $\Et^\gamma$ and \met{} reflects the
detector resolution for unclustered energy.}
\label{fig:pet_vs_met_spectrum}
\end{figure}

Trigger and background considerations drive the choice of the $\Et^\gamma$
threshold. The Level 3 trigger becomes fully 
efficient ($>$ 99\%) at 55 GeV. In addition, 
below 45 GeV the background from \wenu{} with
a misidentified electron 
is very large; as the $\Et^\gamma$ threshold is increased beyond
the kinematic limit for electrons from $W$ decay at rest, the
$W$ must recoil against another object, and the event is then rejected by the
jet and track vetoes.

For an exclusive photon and invisible particle process,
the overall efficiency for all cuts is found to vary from 0.45 at 
\ET{} = 55 GeV to 0.56 for \ET{} $>$ 100 GeV, with a $\pm$10\% uncertainty.
The cumulative effect of each cut is shown in
Fig.~\ref{fig:staged_spectrum}.  The number of events surviving the
photon ID, \met{}, cosmic ray rejection, and jet and track cuts are
15,046, 1,475, 94, and 11, respectively.  The
$\Et^\gamma$ and \met{} in the 11 events in the final sample
are shown in Fig.~\ref{fig:pet_vs_met_spectrum}. 

\begin{table}
\caption{Background sources.  The uncertainty in the QCD background is
unknown, and this background is not considered when setting limits.  The
numbers do not total due to rounding.}
\label{backgrounds}
\begin{ruledtabular}
\begin{tabular}{lc}
Cosmic rays & 6.3 $\pm$ 2.0 \\
$Z\gamma \rightarrow \nu\bar\nu\gamma$ & 3.2 $\pm$ 1.0 \\
$W \rightarrow e\nu$ & 0.9 $\pm$ 0.1 \\
Prompt $\gamma\gamma$ & 0.4 $\pm$ 0.1\\
$W\gamma$ & 0.3 $\pm$ 0.1\\
\hline
Total non-QCD background & 11.0 $\pm$ 2.2 \\
QCD background & $\sim$ 1\\
\hline
Total observed & 11 \\
\end{tabular}
\end{ruledtabular}
\end{table}

To estimate the number of cosmic ray events in the signal sample, we use
the events which have a timing signal
outside the in-time window but which pass all other cuts.  We then
extrapolate into the signal region, assuming a flat distribution in time.

The Monte Carlo simulations of both signal
processes and the
$Z\gamma$, $W\gamma$ and prompt $\gamma\gamma$ backgrounds
use the \textsc{Pythia} event generator \cite{torbjorn} with
the CTEQ5L parton distribution functions (PDFs) \cite{cteq}, 
followed by a parametrized simulation of the CDF detector.
The simulations are then corrected for deficiencies in the detector model
and the $\pm$10\% efficiency uncertainty applied.
We turn off initial state radiation (ISR) to obtain
leading-order (LO) cross-sections and efficiencies.
For the background processes, the resulting cross-sections are corrected 
by the 
ratio of the LO cross-section to the next-to-leading-order 
``zero-jet'' cross-section, obtained from theoretical calculations and Monte 
Carlo estimates.
This allows the correct estimation of the acceptance $\times$ efficiency 
$\times$ cross-section for the exclusive process. 
We obtain correction factors of 0.95 $\pm$ 0.3
for $Z\gamma$ 
\cite{ZgammaKfactor}, 0.9 $\pm$ 0.2 for $W\gamma$
\cite{WgammaKfactor}, and 1.0 $\pm$ 0.3 for prompt $\gamma\gamma$
\cite{ggKfactor};
the systematic uncertainties considered
are $Q^2$ choice and acceptance variations due to modeling of
ISR in the Monte Carlo simulations.  These uncertainties are added in
quadrature with the efficiency uncertainty.

The background from \wenu{} arises either from hard bremsstrahlung by the 
electron before it enters the tracking chamber or inefficiency in the track
reconstruction.  As a radiated photon tends to be collinear with 
the electron,
the \ET{} of the identified electromagnetic object will, in
either case, be close to the initial energy of the electron.
Let $\cal{P}$ be the ratio between the number of electrons
faking photons and the number of electrons passing standard electron 
identification cuts \cite{jeff_prd} in the region $|\eta^e| < 1$; we
estimate $\cal{P}$ 
by assuming that ``$e\gamma$'' events with invariant masses
within 10 GeV of the $\Z$ mass are actually $\Z \rightarrow 
ee$ events.  We obtain $\cal{P}$ $=$ (0.8 $\pm$ 0.1)\%.  The background estimate
is $\cal{P}$ times the number of \wenu{} events that have $|\eta^e| < 1$,
$\Et^e >$ 55 GeV,
\met{} $>$ 45 GeV, and pass the jet and 
track vetoes (discounting the electron track).

We have investigated QCD backgrounds which involve the
mismeasurement of jet energy leading to apparent \met{} or
misidentification of a jet as a photon.  
The most likely
contributors to fakes are events with one high-energy object and many
low-energy jets.
With the \met{}, jet, and track requirements, these events are rare.
To estimate these backgrounds one must use data; however all control
samples have small statistics and estimates range from 0.1 to 2 
events.
We take the conservative approach of not including this background
source in the total background used in the limit calculations.  This can
only make the limits less stringent \cite{omit_QCD}.

We study two hypothetical signal processes in detail.  One is predicted by a 
supersymmetric model and the other by a model with large compact 
extra dimensions.

The first process ($q\bar q \rightarrow \gravitino \gravitino \gamma$) is 
described in \cite{brignole}.  It presumes 
that the gravitino $\gravitino$ is the lightest
supersymmetric particle, with the other superpartners too 
heavy to produce
on-mass-shell at the Tevatron.  Since the gravitino coupling is very small, 
being able to produce
other supersymmetric particles increases the cross-section;
we therefore set an absolute lower limit on the gravitino mass $m_{3/2}$
or, equivalently, the supersymmetry breaking scale $|F|^{1/2}$ (the two are 
related by
$|F| = \sqrt{3} m_{3/2} M_P$, with $M_P$ being the Planck mass).
The cross-section for this process scales as $1/|F|^4$; the kinematic
distributions are independent of $|F|$.

The second process ($q\bar q \rightarrow \gamma G_{KK}$) is described 
in \cite{giudice}:
$n$ extra spatial dimensions are assumed to be compactified with radius $R$.
The fundamental mass scale $M_D$ and $R$ are related
to Newton's constant and the number of extra dimensions by $G_N^{-1} = 
8\pi R^n M_D^{2+n}$ \cite{convnote}.  The standard
model fields propagate only on a 3+1 dimensional subspace, while gravitons
propagate in the whole space.  The graviton modes which propagate in the extra
dimensions appear to four-dimensional observers as massive states of the 
graviton.
A large value of $R$ results in a large phase space for graviton
production, canceling the weakness of the coupling to standard 
model fields.  For a given $n$, the cross-section
scales as $1/M_D^{n+2}$ \cite{efftheory}; for fixed $n$, the kinematic
distributions are independent of $M_D$.

The two signal processes are simulated with modified versions of 
\textsc{Pythia}.  The $q\bar q \rightarrow \gravitino \gravitino\gamma$ 
process is simulated with $|F|^{1/2} = 100$ GeV, and the $q\bar q 
\rightarrow \gamma G_{KK}$ process is simulated with 
$M_D$ = 1 TeV for $n=$ 4, 6, and 8 extra dimensions.

We consider three sources of theoretical systematic uncertainty in the 
cross-section
and acceptance predictions: uncertainty in the choice of $Q^2$ scale,
the choice of parton distribution function, and the modeling of ISR.  
We obtain uncertainty
estimates by varying $Q^2$ by a factor of 4 both up and down,
by using the GRV98 LO PDFs \cite{GRV} instead of the CTEQ5L PDFs,
and by turning the modeling of ISR on and off. 
The uncertainty due to ISR  
includes order-$\alpha_s$ effects and acceptance
changes due to the jet and track vetoes.
For $q\bar q \rightarrow \gravitino \gravitino\gamma$,
the dominant uncertainty is the $Q^2$ choice ($^{+26}_{-18}$\%), followed
by ISR ($\pm$14\%) and PDF choice ($\pm$10\%).
For $q\bar q \rightarrow \gamma G_{KK}$, the dominant uncertainty comes from
ISR ($\pm$34\%), followed by $Q^2$ choice ($^{+18}
_{-16}$\%) and PDF choice ($\pm$8\%).  The overall uncertainty in the
$q\bar q \rightarrow \gravitino \gravitino\gamma$ acceptance $\times$
efficiency $\times$ cross-section, which includes the $\pm$10\% efficiency
uncertainty,
is $^{+33}_{-27}$\%.  For $q\bar q \rightarrow \gamma G_{KK}$,
the corresponding figure is $^{+41}_{-40}$\%.


The method we use to set limits is outlined in \cite{poilim}.
We find the following limits at 95\% C.L.: for the
supersymmetric model, $|F|^{1/2} \ge 221$ GeV (equivalently,
$m_{3/2} \ge 1.17\times 10^{-5}$ eV); for large extra dimensions,
$M_D \ge$ 549, 581, and 602 GeV for $n=$ 4, 6, and 8 extra dimensions
(equivalently, $R \le$ 24 nm, 55 fm, and 2.6 fm, respectively)
\cite{limitsnote}.  The previous best limit published for $|F|^{1/2}$ is 
217 GeV,
from a CDF jet+\met{} search \cite{castro}; the previous best published 
$M_D$ limits set from direct production of
gravitons are 0.68 TeV, 0.51 TeV, and 411 GeV for $n=$ 4, 6, and 8 extra dimensions,
the first two set by DELPHI \cite{delphilimit} and the third by 
L3 \cite{l3limit}.

\begin{figure}
\includegraphics[width=\linewidth]{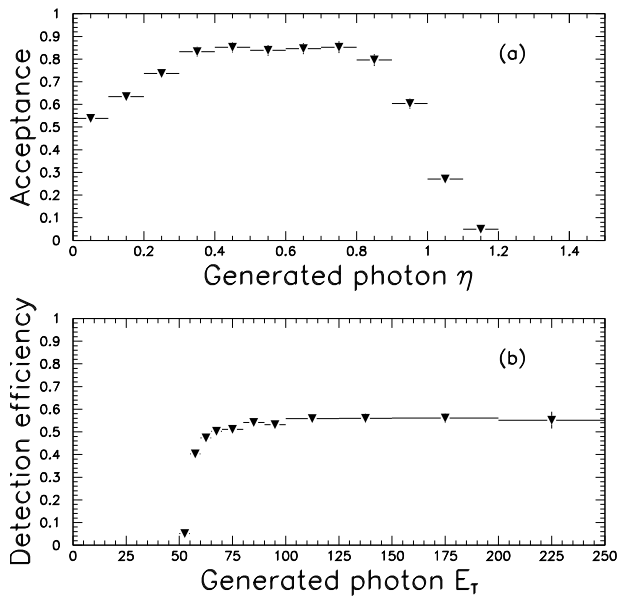}
\caption{Plots of (a) acceptance vs.\ $\eta^\gamma$ and (b) efficiency 
vs.\ $\Et^\gamma$ for the analysis selection.
These plots are valid for any exclusive photon
and invisible particle process.  The error bars are statistical only.
The falloff in acceptance at $|\eta|\simeq 0$ and  $|\eta|\simeq 1$
is due to the folding of the fiducial region of the calorimeter with
the  longitudinal spread ($\sigma \simeq 30$ cm) of the $p\bar p$
collisions.}
\label{fig:modind}
\end{figure}

The results of this analysis can be presented in a `pseudo-model-independent'
manner.  In both the above models,
the uncertainties in the predicted numbers of signal events have been 
dominated by theoretical factors.
It can be useful to derive a limit 
which considers only the uncertainties in the detector simulation
of the processes and so can easily be compared across models \cite{ray_prd}
(keeping in mind that such a limit is not a substitute for the rigorous
extraction of a limit noting theoretical uncertainties).
To obtain this limit, we compute a 95\% C.L.\ upper limit on the number of
events from new physics that would be detected, using only the $\pm$10\% uncertainty
in efficiency as the uncertainty in the acceptance $\times$ efficiency $\times$
cross-section for the new process.
This limit is 9.8 events, which for this integrated luminosity corresponds
to a cross-section of 112 fb.

The plots in Fig.~\ref{fig:modind} allow a comparison of  
models to the acceptance $\times$ efficiency $\times$ 
cross-section limit.  These curves are obtained by
studying the acceptance and efficiency curves for simulated events and
correcting for deficiencies in the detector simulation.  These plots are
valid for both the $\gravitino \gravitino \gamma$ and $\gamma G_{KK}$ processes
studied above, and for any process producing an exclusive photon and 
invisible particle signature.
One can estimate the acceptance $\times$ efficiency $\times$ cross-section for
such a process by convolving the theoretical photon $\eta$ and \ET{} spectra
with the acceptance and efficiency curves.

In conclusion, we have performed a search for new physics in the exclusive
$\gamma$\met{} channel.  We have found no departure from the
expected Standard Model cross-section
and have set limits on two specific models of new physics, one a 
supersymmetric model in which
the photon is produced in association with two gravitinos, the second a
model with large extra dimensions in which the photon is produced in
association with a KK mode of the graviton.
We have also presented the limit in a `pseudo-model-independent' manner.

We thank the Fermilab staff and the technical staffs of the
participating institutions for their vital contributions.  We would
also like to thank J. Lykken, K. Matchev, and D. Rainwater for their
help.  This work was supported by the U.S. Department of Energy and
National Science Foundation; the Italian Istituto Nazionale di Fisica
Nucleare; the Ministry of Education, Culture, Sports, Science, and
Technology of Japan; the Natural Sciences and Engineering Research
Council of Canada; the National Science Council of the Republic of
China; the Swiss National Science Foundation; the A. P. Sloan
Foundation; the Bundesministerium f\"ur Bildung und Forschung,
Germany; the Korea Research Foundation and the Korea Science and
Engineering Foundation (KoSEF); and the Comision Interministerial de
Ciencia y Tecnologia, Spain.

\bibliography{prl}

\begin{thebibliography}{29}
\bibitem[{etd()}]{etdef}
\bibinfo{note}{\ET{} $\equiv$ E $\sin \theta$, where $\theta$ is the polar
  angle of the object measured relative to the event vertex. \met{} is defined
  as the magnitude of the vector sum of \ET{} over all objects in the event.}
\bibitem[{cdf()}]{cdf_coo}
\bibinfo{note}{The CDF coordinate system is right-handed with the $x$ axis
  horizontal and out of the Tevatron ring, the $y$ axis up, and the $z$ axis
  along the proton beam; $\phi$ is the azimuthal angle. The pseudorapidity
  $\eta\equiv-{\rm ln}(\tan (\theta/2))$, where $\theta$ is the polar angle;
  the $\eta$ regions for detector components are defined with respect to the
  center of the detector.}
\bibitem[{eta()}]{eta_note}
\bibinfo{note}{Here $\eta^\gamma$ is defined with respect to the center of the
  detector}.
\bibitem[{uni()}]{units}
\bibinfo{note}{We use the convention that `momentum' refers to $pc$ and `mass'
  to $mc^2$, so that energy, momentum, and mass are all measured in GeV.}
\bibitem[{cor()}]{correctionsnote}
\bibinfo{note}{The \ET{} and \met{} values used in the offline selections are
  corrected values (see Ref. \cite{top_evidence}).}
\bibitem[{omi()}]{omit_QCD}
\bibinfo{note}{Estimates of backgrounds with small central values but large
  uncertainties are necessarily highly asymmetric, as a background cannot be
  less than zero. During the limit-setting procedure, this would produce
  artificially high limits.}
\bibitem[{eff()}]{efftheory}
\bibinfo{note}{We assume that the effective theory is valid at energy scales
  $\lesssim 2M_D$.}
\bibitem[{lim()}]{limitsnote}
\bibinfo{note}{These limits on $M_D$ rise with $n$ because at the points at
  which we can set limits the cross-sections rise with increasing numbers of
  extra dimensions.}
\end{thebibliography}
\end{document}